\documentclass[conference]{IEEEtran}
\IEEEoverridecommandlockouts
\usepackage[utf8]{inputenc} %unicode support
\usepackage{cite}
\usepackage{amsthm,amsmath,amssymb,amsfonts}
\usepackage{algorithm}
\usepackage{algpseudocode}
\usepackage{graphicx}
\usepackage{textcomp}
\usepackage{xcolor}
\usepackage{listings,multicol}
\usepackage{enumitem}
\usepackage{float}
\usepackage{subfig}
\usepackage{url}

\newtheorem{definition}{Definition}
\newcommand{\etal}{et al.}

\def\BibTeX{{\rm B\kern-.05em{\sc i\kern-.025em b}\kern-.08em
    T\kern-.1667em\lower.7ex\hbox{E}\kern-.125emX}}
\begin{document}

\title{Energy Smart Buildings: Parallel Uniform Cost-Search with Energy Storage and Generation
\thanks{The presented research is partially funded by the Dutch Research Council (NWO) in the framework of the Indo-Dutch Science Industry Collaboration programme with project NextGenSmart DC (629.002.102).}
}

\author{\IEEEauthorblockN{Brian Setz\IEEEauthorrefmark{1},
		Kawsar Haghshenas\IEEEauthorrefmark{2}, and Marco Aiello\IEEEauthorrefmark{3}}
	\IEEEauthorblockA{Service Computing Department,
		University of Stuttgart\\
		Stuttgart, Germany\\
		Email: \{firstname\}.\{lastname\}@iaas.uni-stuttgart.de}
}
%
%\author{\IEEEauthorblockN{Brian Setz}
%\and
%\and
%\and
%\and
%\and

\maketitle

\begin{abstract}
The amalgamation of Internet of Things and the smart grid enables the energy optimal scheduling of appliances based on user needs and dynamic energy prices. Additionally, progress in local storage technology calls for exploiting additional sources of flexibility. In this paper, we propose a scheduling approach for building operation management, considering factors such as energy storage, local energy generation, and dynamic energy prices. In addition, we propose a new optimization strategy to discover the optimal scheduling of devices. Our approach utilizes parallel uniform cost-search to explore the complex search space and to find the optimal schedule within a user-acceptable amount of time. The evaluation utilizes real-world data for the devices, and the price signals, while the architecture is designed following a micro-service approach, enabling modularity and loose-coupling. The evaluation shows that including local energy storage as part of the optimization problem further reduces overall costs by up to 22.64\% when compared to schedules without energy storage. Parallel uniform cost-search decreases the time to find the optimal schedule by a factor of 4.7 with respect to the traditional uniform cost-search algorithm.
\end{abstract}


\section{Introduction}

In 2022, gas and electricity prices have surged in Europe as fossil fuel prices rise, emphasising the need for solutions  to reduce energy use~\cite{Ari2022}. Buildings are a major source of energy usage and a natural place to look for cutting consumptions and costs. Building operations management encompasses the controlling and monitoring of buildings, and can assist in the optimization of buildings. Building management benefits greatly from two emerging paradigms: the smart power grid and the digital interface to everyday objects known as the Internet of Things (IoT). The smart grid aims to integrate the behaviours and actions of all connected actors, be it consumers, producers, or prosumers by leveraging on information and communication technologies. Local energy generation and storage, economic efficiency, and sustainability are key concepts associated with the smart grid~\cite{TeixeiraMartins2019}. Whereas the IoT paradigm enables a wide variety of devices and machines to interact with one another programmatically and autonomously~\cite{Lee2015}. This enables the vision of smart buildings, which are able to use devices to sense and control, and in turn influence the energy consumption of the building by reacting to different signals while satisfying user needs. The dynamic real time price of energy is a prototypical signal which the smart grid is envisioned to provide to end users~\cite{Chen2009}. This signal is influenced by weather condition, fossil fuel prices, and current demand~\cite{Wang2012,6832441}. End users receive an incentive to actively participate by changing and rescheduling their load patterns based on current prices, while not being forced to~\cite{Gholian2016,Fiorini2019}. This differs from demand response schemes where the users leave the ultimate control of their consumption to the energy provider~\cite{Darby2013}.

Smart buildings are able to sense their context of use and control systems and devices based on user requirements and external signals~\cite{Nizamic2014/08}. As a reaction to the dynamic price signal, the building can operate on subsystems by actuating on them, and by postponing their operation to times where the signal is expected to be more favourable. For example, the scheduled time of heating for a meeting room can be advanced or postponed, if the energy prices are expected to increase considerably during the scheduled heating period. However, deciding when to schedule a load in order to minimise costs is a non trivial task, especially when local energy generation and storage add uncertainty to the overall set of possible decisions to make. 

We have previously proposed an architecture and optimization strategy to control office appliances based on user needs and dynamic prices~\cite{Georgievski2012}. Since then, we have witnessed a widespread adoption of IoT solutions to monitor and control appliances and building systems, IoT and data science approaches to building context and human activity recognition, as well as rapid developments in electricity storage technologies. This provides both challenges and opportunities for building energy optimization. 

One of the major challenges are that more uncertainty is present in the system and that the complexity of monitoring and decision making increases. The opportunity comes from the major flexibility and controllability of the system. Therefore, we approach the problem considering the current state of the art and resort to improved techniques for the optimal scheduling of office buildings. Specifically, we present an approach to find the optimal load schedule that minimises costs while considering (1) price signals from the grid and prosumers, (2) local energy generation, (3) local energy storage, and (4) the scheduling constraints of typical office devices. Based on this information, an optimization problem is formulated. The resulting search space is large, as for each time slot there are many permutations of possible actions that can be taken. The contributions made and presented here include:
\begin{itemize}
	\item A general optimal device scheduling algorithm for smart grid enabled buildings that considers: photovoltaic panels, micro wind turbines, and battery energy storage. Our evaluation shows a reduction in costs by up to 22.64\%;
	\item A parallel uniform-cost search algorithm to find the optimal schedule that minimizes overall costs while ensuring performance when navigating complex search spaces. We observe a decrease in search time by a factor of 4.7 compared to traditional uniform-cost search;
	\item A micro-service architecture for the integration of building management systems with independent cloud services, enabling a loosely coupled, modular systems that can be adjusted for any building; and,
	\item An evaluation using real data for the scheduled devices and for the prediction of renewable energy generation. The device data was collected from an office environment over a period of eight months through plug load monitoring, with measurements taken every 10 seconds. 
\end{itemize}

The rest of the paper is organised as follows. In Section~\ref{sec:relwork}, we review related work and place the present contribution in the context of it. The problem definition, proposed approach, models, scheduling policies, and the architecture of the various components of the systems are described in Section~\ref{sec:approach}. This is followed by the evaluation with real data, in Section~\ref{sec:eval}. The evaluation considers both the economic and performance aspects. The discussion of the results, the limitations, as well as concluding remarks are summarized in Section~\ref{sec:discuss}.
\section{Related Work}
\label{sec:relwork}

A common approach to the scheduling of loads is using Mixed-Integer Linear Programming (MILP) models. Paterakis~\etal~present a method for optimal scheduling of household appliances~\cite{Paterakis2015}. The scheduling is performed under hourly pricing and peak power limiting demand response strategies. The authors utilize a MILP model for optimal scheduling, distinguishing between thermostatically and non-thermostatically controllable devices. The results show that the load factor can be improved significantly, while the economic costs slightly increase. Duman~\etal~also demonstrate the use of a MILP model for scheduling the operation of shiftable loads~\cite{Duman2021}. The distinction is made between time-shiftable loads, power-shiftable loads, and thermostatically controlled loads. Their simulation shows that daily costs can be reduced by 53.2\%. A similar method has been proposed by Nan~\etal~for optimal demand response scheduling in residential communities~\cite{Nan2018}. A clear distinction is made between interruptible loads, shiftable loads, and distributed generation. The authors define a MILP problem to be solved in order to find a solution that reduces peak load and peak-valley differences and, in turn, the overall cost of power. The end user costs are reduced by up to 1.52\%. 

While MILP appears to be the predominant approach, alternatives have been proposed. Amer~\etal~ propose a home energy management system centred around a multi-objective optimization problem~\cite{Amer2021}. A distinction is made between shiftable loads, non-shiftable loads, and active loads. Energy storage and PV energy resources are also considered. The objective is to balance the benefits for the end-user and the Distribution System Operator (DSO). The results show a reduction in energy cost of 31\%, and a reduction in demand peak by 18\%. Lu~\etal~present a different approach for demand response management using reinforcement learning and artificial neural networks~\cite{Lu2019}. While a distinction between different types of loads is made, energy storage and renewable generation are not considered. The obtained cost savings are between 7.3\% and 72.3\%. The authors demonstrate that their approach outperforms MILP as the number of iterations increases. The method proposed by Li~\etal~utilises deep-reinforcement learning for the optimal scheduling of home appliances~\cite{Li2020}. The appliances are split into three categories: deferrable devices, regulatable devices, and critical devices. The approach reduces the overall electricity costs by 31.6\%. An approach that is more closely related to the work that is presented in this paper, is the work of Fioretto~\etal~\cite{Fioretto2017}. They define a constraint optimization problem, which is solved by a distributed algorithm that divides the problem into individual sub-problems to overcome the complexity of the problem. The savings compared to the baseline use case are over 50\%. However, in our work the problem is not divided into sub-problems, as this would lead to non-optimal global solutions.

Finally, in our previous work, we demonstrated that the optimal scheduling of the operation of devices based on user-defined policies can result into significant savings~\cite{Georgievski2012}. From the economic perspective, the obtained savings are 35\% on average. Furthermore, energy savings of 10\% are also observed. The occupant satisfaction study in our previous work shows that comfort and satisfaction is preserved while our system operates the devices. In this work, we further tackle more complex scheduling problems by including real-time predictions of renewable energy generation, and the possibility for energy storage. To counteract the increased complexity, a novel algorithm is proposed that parallelises the solution of to the scheduling problem. The distinct subsystems are encapsulated in a micro-service architecture that promotes modularity and scalability.

The approach taken in our work also recognises the categorizations of shiftable, non-shiftable, and active loads as done in several existing studies. The difference is that these properties are encapsulated in fine grained policies which are assigned to individual devices. We consider energy storage to be a property of a device, and therefore it is also an assignable policy. Furthermore, we guarantee to find the optimal solution to the scheduling problem, as all possible and valid permutations of the state space are explored. In addition, the approach proposed in this work is distributed in nature, both the novel parallel uniform-cost search algorithm and the underlying micro-service architecture can be distributed across multiple machines. Additionally, we use real-world and real-time data from multiple sources to predict the renewable energy generation and model the day-ahead prices. To the best of our knowledge, our proposed approach is the first to combine these aspects into a singular approach.

\section{Approach}\label{sec:approach}
First we define the energy optimization problem in terms of the
relationships between devices, their desired behaviour, and the
available energy sources. This is followed by the models of the energy
sources, in the present case: wind turbines, photovoltaic panels, and
prosumers on the same local grid. Energy storage in the form of
batteries is also modelled as part of the optimization
problem. The last part of the problem definition relates to the user
policies associated to the office devices. Such policies drive the way
in which each one can be scheduled. Finally, we describe the algorithm to 
find the optimal schedule, and provide an overview of the proposed architecture.

\subsection{Problem Definition}

Given the devices to be scheduled, the scheduling policy for every
device, and the available energy sources, the goal is to find an
optimal day-ahead schedule that minimizes the costs while ensuring
all of the constraints as defined by the scheduling policies are
satisfied. The following problem definition is based and extends the
definitions from our previous work~\cite{Georgievski2012}.

A day is divided into time slots with a fixed duration. For every time slot $t$ with length $t.duration$, $ES(t) = \{ es_i \}$ represents the available energy sources at time slot $t$. An energy source consists of a tuple $es_i = \langle cost, energy \rangle$, where cost is the price per kWh at time $t$ for energy source $i$, and energy is the available energy that the given energy source can provide at time $t$. The energy sources include both renewable energy sources and neighbouring prosumers.

The set of all building devices that are amenable to scheduling is
$D$. For each device $d_i \in D$ there exists a tuple
$d_i = \langle d_{id}, S_i \rangle$ where $d_{id}$ is the identifier
of the device, and $S_i$ is the set of available states in which the
device can be. Every state $s_{ij} \in S$ consists of the tuple
$\langle s_{id}, power \rangle$ where $s_{id}$ is the identifier of
the state, and $power$ is the power consumption in that particular
state.

Each device $d_i$ has associated one or more scheduling policies that
determine the scheduling behaviour for the device. 
The set $P$ contains all of the policies that are associated with devices. Each policy $p_i \in P$ is represented by the triple
$p_i = \langle p_{id}, type, parameters \rangle$ where $p_{id}$ is the
identifier of the policy, $type$ is the policy type, and $parameters$ are
the specific parameters for that type of policy.

\begin{figure}[htbp]
	\centering
	\includegraphics[width=1.0\columnwidth]{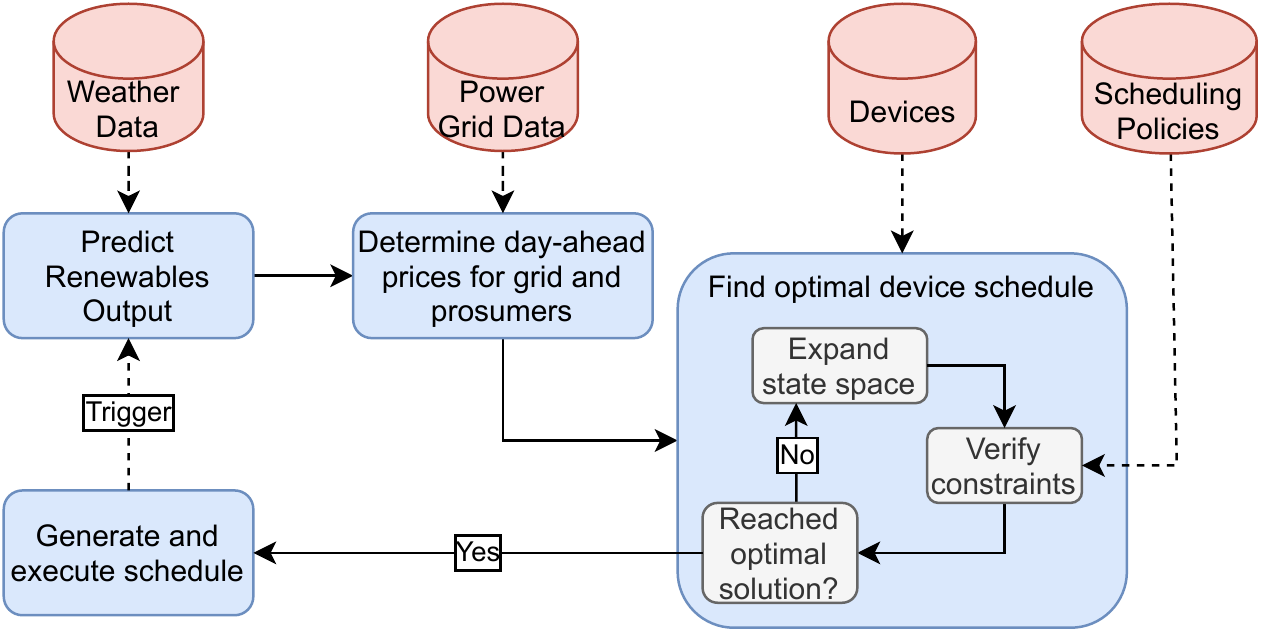}
	\caption{A high-level overview of the approach proposed in this work}
	\label{fig:overview}
\end{figure}

A schedule $X = \{ x_{ti} \}$ is a set of values per device per time
slot, where each value $x_{ti} \in S_i$ denotes the state that a
device $i$ assumes at time $t$.  Then we have:
\begin{definition}[Office optimization problem] Find a schedule $X =
  \{ x_{ti} \}, \forall t \in T, \forall i \in D$ that is optimal. An
  schedule is optimal iff
\begin{equation}
\sum_{t \in T} cost(t, e_t) \to min, \forall p \in P : satisfied(p, X)
\end{equation}
where
\begin{equation}
e_t = \sum_{d_i \in D} x_{ti}.power * t.duration
\end{equation}
\end{definition}
In other words, the schedule is optimal when the devices are scheduled
in such a way that the policy constraints for each device are
satisfied and the total costs of operation are minimal. To find the optimal solution requires exploration of the state space. The state space consists of all possible permutations of device states for each time slot. A schematic overview of the approach we propose is given in Figure \ref{fig:overview}.

The process of finding an optimal schedule is started by a trigger, for example, the beginning of a new day. First, the energy generation of the renewable energy sources is estimated based on weather data and individual models for each renewable energy source. Next, the cost of energy is determined for each of the prosumers as well as the DSO. All of this data is passed to our scheduling algorithm which continuously expands the state space, verifies constraints, and checks if the optimal solution is found. The complexity of this state space depends on the set of devices, the assigned scheduling policy for each device, and the scheduling horizon. Once the optimal solution is found, a schedule is generated in JSON format, which can then be executed by the building management system. The process is repeated whenever a trigger takes place.  

\subsection{Energy Sources: Wind Turbines}

One of the energy sources that is considered is local (on-site) energy
generation using wind turbines. The following mathematical model is
used to represent the wind turbine's power output and is based on the
work of Xia \etal~\cite{Xia2013}:

\begin{equation}
	P_{w} = \frac{1}{2} \rho C_p A v^3_w
\end{equation}

where:
\begin{description}[labelsep=3.5em, align=left, labelwidth=3.5em,labelindent=2em]
	\item [$P_{w}$] power output (W);
	\item [$\rho$] air density (kg/m³);
	\item [$C_p$] power coefficient;
	\item [$A$] wind turbine swept area (m²);
	\item [$v_w$] wind speed (m/s).
\end{description}
\vspace{1em}
The wind turbine swept area and the power coefficient are dependent on
the particular wind turbine model. As per the Betz Limit, the
theoretical maximum power coefficient is 0.59, though in practice the
power coefficient is generally in the 0.35-0.45
range~\cite{RAE2010}.

To determine the air density, first the saturation vapour pressure must
be calculated. We resort to the work of Herman Wobus who has provided
an approximating polynomial to calculate the vapour
pressure~\cite{CurtisGerald2003}:

\begin{equation}
	\begin{split}
	p_s &= e_{so} / C^8 \\
	C &= c_0 + T_d (c_1 + T_d (c_2 + T_d(c_3 + T_d(c_4 + \\ 
	&T_d(c_5 + T_d(c_6 + T_d (c_7 + T_d (c_8 + T_d c9))))))))
	\end{split}
\end{equation}

where:
\begin{description}[labelsep=3.5em, align=left, labelwidth=3.5em,labelindent=2em]
	\item [$p_s$] saturation vapour pressure (hPa);
	\item [$e_{so}$] known constant (6.1078);
	\item [$c_0-c_9$] known constants;
	\item [$T_d$] dew point (\textdegree{}C).
\end{description}
\vspace{1em}
Once the saturation vapour pressure is known, the air density can be
calculated according to the following equation:
\begin{equation}
	\begin{split}
	p_d &= p - p_s \\
	\rho &= \frac{100 p_d}{R_{da} T_{ok}} + \frac{100 p_s}{R_{wv} T_{ok}}
\end{split}
\end{equation}
where:
\begin{description}[labelsep=3.5em, align=left, labelwidth=3.5em,labelindent=2em]
	\item [$p_d$] dry air pressure (hPa);
	\item [$p$] air pressure (hPa);
	\item [$p_s$] saturation vapour pressure (hPa);
	\item [$\rho$] air density (kg/m³);
	\item [$R_{da}$] dry air gas constant (J/kgK);
	\item [$R_{wv}$] water vapour gas constant (J/kgK);
	\item [$T_{ok}$] outside air temperature (K).
\end{description}
\vspace{1em}
And finally, the cut-in and cut-out air speeds of the wind turbine
have to be taken into account. If the air speed is above (cut-out) or
below (cut-in) a certain value, the output of the turbine will be
zero:
\begin{equation}
	P_{turbine} =
	\begin{cases}
		P_w \quad &\text{if} \, v_{in} \leq v_w \leq v_{out}\\
		0 \quad & \text{otherwise} \\
	\end{cases}
\end{equation}
where:
\begin{description}[labelsep=3.5em, align=left, labelwidth=3.5em,labelindent=2em]
	\item [$P_{turbine}$] turbine power output (W);
	\item [$v_{in}$] cut-in air speed (m/s);
	\item [$v_{out}$] cut-out air speed (m/s).
\end{description}
\vspace{1em}
The cut-in and cut-out air speed are dependent on the wind turbine model.

\subsection{Energy Sources: Photovoltaic Panels}

Photovoltaic panels are a common local generation means. We use the
following formulation to model the photovoltaic panel's power output
based on the work of Chen\etal~\cite{Chen2012}:
\begin{equation}
	P_{photovoltaic} = \eta SI(1 - 0.005 (T_{oc} - 25))
\end{equation}
where:
\begin{description}[labelsep=3.5em, align=left, labelwidth=5em,labelindent=2em]
	\item [$P_{photovoltaic}$] photovoltaic power output. (W);
	\item [$\eta$] panel efficiency (\%);
	\item [$S$] array area (m²);
	\item [$I$] solar irradiation (W/m²);
	\item [$T_{oc}$] outside air temperature (\textdegree{}C).
\end{description}
\vspace{1em}
The array area and panel efficiency depend on the configuration and on
the model of the photovoltaic panels. The solar irradiation and
outside air temperature depend on the location and weather. This model
assumes that the panels utilize a two-axis maximum power point
tracker. As a result, the surface of the panel is always perpendicular
to the sunlight. Therefore, the direct normal solar irradiance is used
in the calculation of the panel's power output.

\subsection{Energy Sources: Prosumers and DSO}

The final energy sources considered in this work are the energy
provided by neighbouring prosumers and by the DSO. These are both seen as energy providers
independently of the underlying technology used. For instance the
prosumer could also have photovoltaic panels or storage systems, while
the DSO also has access to the energy coming from the
transmission infrastructure. For both, we need to define the cost of the
energy, and the amount of energy available during a given time
slot. For prosumers, the available energy (kWh) is sampled from a normal
distribution:
\begin{equation}\label{eq:uniformdis}	
	X \sim \mathcal{U}_{[a, b]}\,
\end{equation}
whereas the cost of the energy (€/kWh) is sampled from a Gaussian
distribution:
\begin{equation}\label{eq:gaussiandiss}
	X \sim \mathcal{N}(\mu,\,\sigma^{2})\,
\end{equation}

As typical in practice, we assume that the DSO can always fulfil the
demand of the system. Though, the cost of energy from the DSO will 
typically be higher than those of prosumers, especially in times of
high demand. As for the cost of energy supplied by the DSO, this is modelled 
after the day-ahead energy market prices for DSOs, which are then scaled to
values more representative of real-world end-user prices. 

\subsection{Energy Storage: Battery}

Energy storage provides for important flexibility in complex smart
energy systems. It can take many shapes and sizes, from fly wheels, to
rechargeable batteries, to hydroelectric dams. In this work,
residential energy storage in the form of household batteries is
considered as it is the most easy and likely to be deployed in an
office building. The battery should adhere to the following
constraints, based on the definitions by Chen
\etal~\cite{Chen2012}. If $C(t)$ is the charge of the battery at time
$t$, $P^c_t$ the power charged to the battery, $P^d_t$ the power
discharged from the battery, and $\eta$ is the charging
efficiency, then
\begin{equation}
	\begin{split}
	\text{Charge:}&\; C(t+1) = C(t)  + \frac{\Delta t P^c_t}{\eta} \\
	\text{Discharge:}&\; C(t+1) = C(t)  - \frac{\Delta t P^d_t}{\eta}
	\end{split}
\end{equation}
under the maximum charge and discharge limits:\\
\begin{equation}
	\begin{split}
		0 \leq P^c_t \leq P^{c,max}_t \\
		0 \leq P^c_t \leq P^{c,max}_t \\
	\end{split}
\end{equation}
energy storage limits:\\
\begin{equation}
	C_{min} \leq C(t) \leq C_{max}
\end{equation}
and the start and end limits:\\
\begin{equation}
	C(t_{start}) = C(t) = C(t_{end})
\end{equation}
where $P^{c,max}_t$ and $P^{d,max}_t$ are the maximum charge rate and
maximum discharge rate respectively, $C_{min}$ and $C_{max}$ are the
minimum and maximum energy that the battery can store, and
$C(t_{start})$ and $C(t_{end})$ represent the charge of the battery at
the beginning and end of the scheduling period.

\subsection{Scheduling Policies}

The way devices operate varies depending on their main function and
user needs. This influences the way they are scheduled. To represent
this, we have introduced several policies to represent
usage~\cite{Georgievski2012}. The policies are then translated into
scheduling constraints when solving the optimization problem. The following scheduling policies are available:

\begin{itemize}
\item \textbf{Total Policy}: defines the total amount of time a device
  $d_i$ should be in state $s_{ij}$. For example, for any device with
  a battery that needs to be recharged, this is the time it takes to
  fully charge it. The amount of time is fixed, though it need not be
  continuous.
	
\item \textbf{Continuous Policy}: similar to the \textit{total}
  policy, if defines the total amount of time a device $d_i$ should be in
  state $s_{ij}$. The key difference being that once the device is in
  state $s_{ij}$, it should remain in this state for the given amount
  of time without interruptions.
	
\item \textbf{Repeat Policy}: defines the amount of time a device
  $d_i$ should be in state $s_{ij}$ with a certain periodicity. For
  example, a freezer that should be scheduled repeatedly every hour to
  maintain the appropriate temperature.
	
\item \textbf{Multiple Policy}: combines the \textit{continuous} and
  \textit{repeat} policies. This policy is for devices that need to be
  in state $s_{ij}$ for a certain number of times, but should not be
  interrupted before a certain deadline once they enter that
  state. For example, a printer can have multiple jobs to complete,
  but cannot be interrupted while executing the job.
	
\item \textbf{Strict Policy}: defines a schedule that is known ahead
  of time, determining exactly when device $d_i$ should be in state
  $s_{ij}$, for every available time slot. For example, a safety light
  that should be on from dusk till dawn. It would not be acceptable to
  only turn the light on for 15 minutes every hour.
	
\item \textbf{Pattern Policy}: similar to the \textit{strict} policy, it
  defines ahead of time what the expected usage of a device
  is. However, this policy does not offer the possibility of the
  device to be controlled, it simply provides the expected usage
  pattern.
	
\item \textbf{Sleep Policy}: defines a no-op policy, during the entire
  specified period, device $d_i$ should be in state
  $s_{ij}$ and should not be operated upon. For example, this is useful
  when a device should be turned off for a period of time in order to
  save energy.
	
\item \textbf{Battery Policy}: indicates that a device $d_i$ is able
  to act as short-term energy storage, allowing for charging and
  discharging of energy. Examples of devices that can have the battery
  policy include stationary home energy storage devices, as well as connected
  electric vehicles.
\end{itemize}

The formal definitions for each policy, except for the
\textit{battery} policy which we introduce for the purpose of the
present work, can be found in~\cite{Degeler2014}. 

\subsection{Parallel Uniform-cost Search}

The goal of this work is to find the optimal solution to the previously defined problem. Optimal in this case, refers to the solution that minimizes
the overall costs. The cost of the solution can be defined in multiple ways, for example, as the total cost of electricity, or as the carbon footprint associated with the energy consumption. To solve the problem optimally, the state space has to be explored. The state space can be interpreted as a weighted graph, where each node is a permutation of the state space, edges represent the  transformation from one state to another and have an associated cost, the depth of the graph is equal to the number of timeslots. Uniform-cost search is an uninformed search algorithm used 
for traversing graphs, finding the path with the lowest cost from the root node to the destination node \cite{Russell2010}. This work presents a novel parallel uniform-cost search algorithm, which finds the optimal device schedule significantly faster than regular uniform-cost search.

The proposed algorithm is given in Algorithm \ref{alg:ucsp}. The search is initialized by creating an empty node and adding it to the global priority queue. Each node contains a permutation of device states, and the cost associated with being in this state. The priority queue ensures that the node with the lowest cost is the first node in the queue. Each thread also maintains a local priority queue. When the local queue is empty, the thread takes a node from the global queue and starts expanding the state space while populating its local queue. These newly expanded states are added back into the priority queue, assuming none of the policies are violated and they pass the heuristic checks. This is repeated until one of the threads finds a solution to the scheduling problem. Once a possible solution is found, the local queues are merged with the global queue to verify that there are no nodes with a lower cost in any of the local queues. If this is true, then the optimal solution is returned.

Two heuristics are applied to reduce the complexity of the search space. The first heuristic check is to determine whether the newly expanded
states violate any policies. If a policy is violated, the state is invalid and should not be explored further. For example, a device with a \textit{sleep} policy from 01:00 AM till 06:00 AM should not be scheduled during this given period. If the device is scheduled during this period in the newly expanded state, then the state violates the policy and therefore the child node is pruned from the search space. The second heuristic checks if there are partial solutions that have already been explored and that have the same outcome (cost and number of time slots) as the newly expanded state. If the outcome of the partial solution is a duplicate, it is discarded. In case the outcome is the same, but the cost is lower, the original node with the same outcome is replaced by the newly created child node. In case the outcome of both nodes are identical, the child node is pruned from the search space.

\begin{algorithm}[ht]
	\caption{Parallel Uniform-cost Search}
	\label{alg:ucsp}
	\begin{algorithmic}[1] 
		\Procedure{UCS\_Parallel}{problem, threads} 
		\State initialNode $\gets$ new(node)
		\State globalQueue $\gets$ push(globalQueue, initialNode)
		\State globalVisited $\gets \emptyset$
		\State solutionFound $\gets$ False
		\For{i $\gets$ 1 to threads} 
		\State{startNode $\gets$ pop(globalQueue)}
		\State{localQueue $\gets$ push(localQueue, startNode)}
		\State{localVisited $\gets \emptyset$}
		\While{\textbf{not} solutionFound}
		\State{node $\gets$ pop(localQueue)}
		\If{goalReached(problem, node.state)}
		\State{synchronizeThreads(threads)}
		\If{peek(globalQueue).cost $>$ node.cost}
		\State{solutionFound $\gets$ True}
		\State{\textbf{return} solution(node)}
		\EndIf
		\EndIf
		\State localVisited $\gets$ insert(localVisited, node)
		\For{newState \textbf{in} expand(problem, node.state)}
		\State c $\gets$ createChild(problem, node, newState)
		\If{policiesViolated(c.state)} 
		\State \textbf{continue} 
		\ElsIf{c.state \textbf{not in} localQueue \textbf{or} visisted} 
		\State localQueue $\gets$ push(localQueue, c)
		\ElsIf{c.state \textbf{in} localQueue \textbf{with} higher cost}
		\State localQueue $\gets$ replace(localQueue, c.state, c)
		\Else
		\State \textbf{continue}
		\EndIf
		\EndFor
		\If{length(localQueue) $\leq$ threshold}
		\State{Merged localQueue into globalQueue}
		\State{Merged localVisited into globalVisited}
		\EndIf
		\EndWhile
		\EndFor
		\State \textbf{failure}
		\EndProcedure
	\end{algorithmic}
\end{algorithm}

\subsection{Architecture}\label{sec:arch}

\begin{figure*}[htbp]
	\centering
	\includegraphics[width=0.75\textwidth]{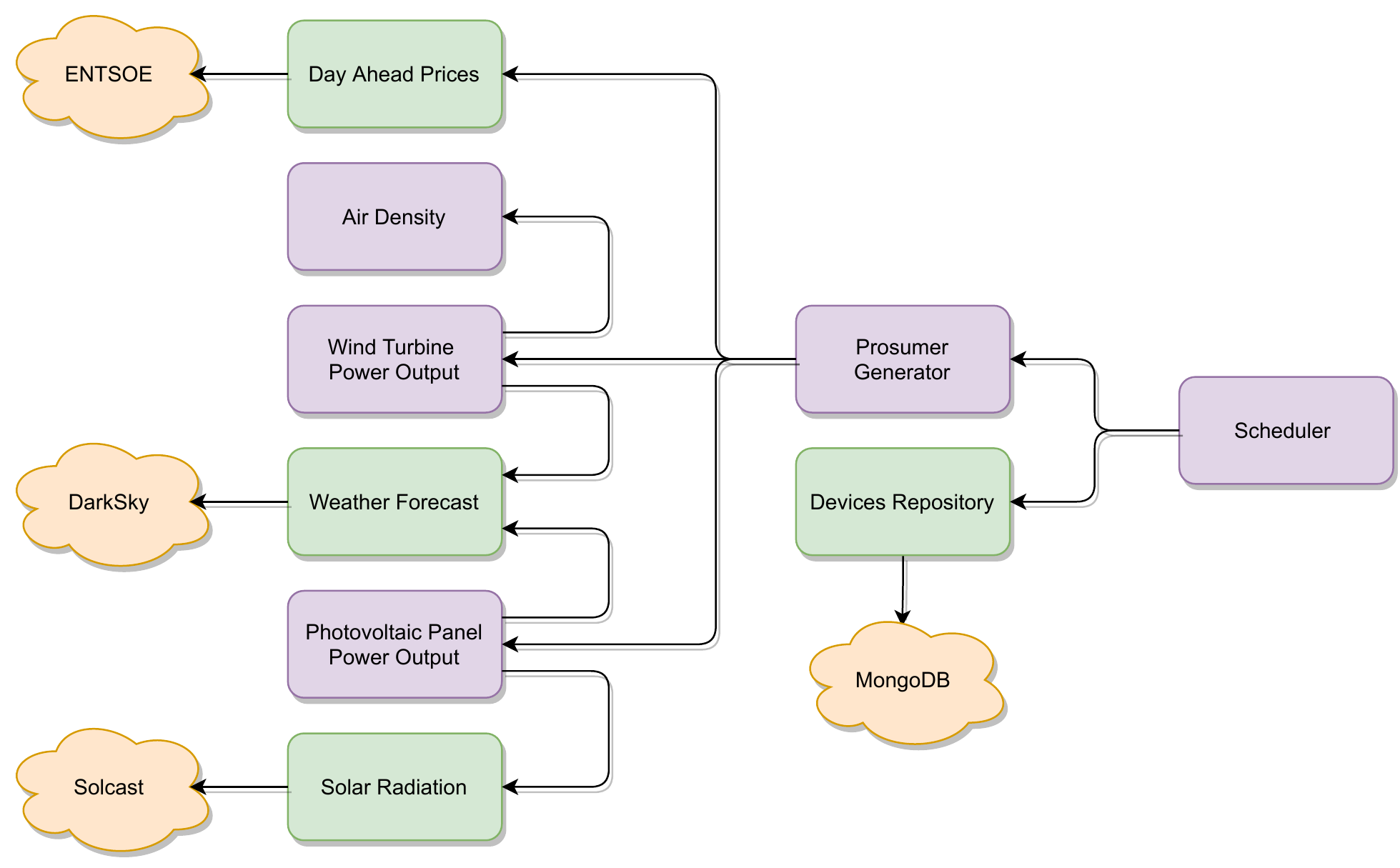}
	\caption{Microservice architecture. The green coloured shapes represent services that can easily be exchanged to support a different data source. The purple shapes are the core services. The orange shapes are external services. The arrows represent the interdependencies.}
	\label{fig:microservices}
\end{figure*}

The cost optimization of smart grid enabled buildings requires the availability of a wide variety of data sources. For example, the reliable prediction of the renewable generation for the next day is based on the weather forecast. Weather services vary considerably in terms of the predicted features (radiation, temperature, wind, humidity, etc.), the spatial granularity, and the temporal resolution. The microservice architecture that we design and propose enables the decoupling of the system's components, where each loosely coupled service adheres to the single responsibility principle~\cite{Bakshi2017}. Furthermore, this architectural design promotes modularity, allowing services to be easily replaced by alternative services, even at runtime~\cite{Kalske2018}. An example is replacing the weather forecast of provider A, by the forecasting service of provider B, because provider B provides better local forecasts at an increased resolution without affecting the overall operation of the system.
The proposed microservice architecture is shown in Figure~\ref{fig:microservices}. It consists of nine distinct microservices, identified by the rounded rectangles. The arrows indicate the interdependencies between them. Four of these services depend on external components, which can be cloud services (e.g. DarkSky for weather data), or data repositories (e.g. MongoDB for storing device data). In case a different weather forecast supplier should be used, it is a matter of replacing the weather forecast service with a different one. Communication between services is exclusively done via the REST architectural style, as each REST service exposes a REST API that accepts HTTP requests. The Content-Type of each HTTP response is \textit{application/json}. The deployment of the microservices is handled by means of containerization, where each service is hosted within individual Docker container. What follows next is a description of each microservice.

\begin{description}
	\item[Day Ahead Prices Service:] requires an Energy Identification Code to identify the energy market that is used, and a date for which the day ahead prices should be retrieved. This service uses the cloud API provided by the European Network of Transmission System Operators transparency platform \cite{Entsoe} in order to obtain realistic and real-time day ahead prices. The services returns the hourly day ahead prices for the specified date.
	
	\item[Air Density Service:] requires the temperature, air pressure, and dew point in order to calculate the air density. This service implements the Herman Wobus vapour pressure polynomial to calculate the saturation vapour pressure at the dew point temperature, which can then be used to calculate the pressure of dry air. Finally, the service returns the air density for the given parameters.
	
	\item[Wind Turbine Power Output Service:] requires the turbine's blade radius, wind speed, air density, and power coefficient to calculate the turbine's power output. This service also requires the cut-in and cut-out wind speed of the turbine to be provided. Based on these parameters, the turbine power output is given.
	
	\item[Weather Forecast Service:] requires the latitude and longitude of the location for which the weather forecast should be obtained. This service returns the temperature, dew point, humidity, air pressure, and wind speed. The weather service used in this work is DarkSky~\cite{DarkSky} as it provides high temporal resolution forecasts for our desired location.
	
	\item[Photovoltaic Panel Power Output Service:] requires the direct normal solar irradiance, temperature, area of the panel, and the efficiency of the panel. The service assumes that a maximum power point tracker is used. Based on the parameters provided to the service, the expected power output is returned. 
	
	\item[Solar Radiation Service:] requires the latitude and longitude of the location for which the solar radiation should be collected. The solar irradiance data is provided by the cloud API from Solcast~\cite{Solcast}. This service returns the direct normal irradiance, diffused horizontal irradiance, and global horizontal irradiance. 

	\item[Prosumer Generator Service:] requires the location (latitude and longitude) for which the data for simulated prosumers should be generated, as well as the specifications of the available PV and/or wind turbine resources. This service generates a list of energy sources which includes prosumers representing the neighbouring buildings connected to the smart grid. Each prosumer is able to deliver a certain amount of energy for a certain cost during a certain period of time. Added to this list of prosumers are the available renewable resources of the building in question, as well as an energy source representing the (traditional) grid in case the energy provided by the renewables and prosumers is insufficient to satisfy the demand.
	
	\item[Device Repository Service:] returns the list of devices which should be scheduled, their power consumption in different states, as well as the policy for each device. The devices, their power states and their policies are stored in the document-oriented database MongoDB. 
	
	\item[Scheduler Service:] requires a list of energy sources (prosumers), a list of devices to schedule, the initial charge of the battery, as well as the algorithm that should be used. The modularity allows for selecting among versions of the uniform grid-search problem algorithm. The service returns the day ahead schedule of the devices with a 15 minute granularity, this predetermines the power state that each of the devices should be in for every 15 minute interval. Some meta-data on the performance of the algorithm is also included in the data that is returned by this service.  
\end{description}
\section{Evaluation}
\label{sec:eval}

To evaluate the quality and possibilities of the proposed architecture and scheduling approach, we consider offices. The specific setup consists of individual offices and a shared kitchen area, each with numerous devices. Smart plugs are used to collect power data from the devices situated in the offices and the kitchen area. This data is used to define the devices' profiles required by the scheduler. The renewables energy sources and energy storage are modelled after commercially available products. Combined with real-world weather data, the power output of the renewables is computed. The device profiles and energy sources are then used as input for the scheduler. Three different cases are considered; for each case we evaluate it with and without the availability of energy storage. For each generated schedule, the economic cost, the energy demand profile, and the battery charge are analysed.

\begin{figure}[!t]
	\centering
	\subfloat[Histogram of power data \label{fig:screenHistogram}]{\includegraphics[width=1\linewidth]{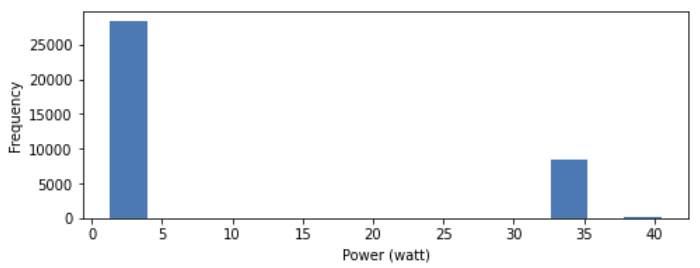}}\\
	\subfloat[Clustering of power data \label{fig:screenClustering}]{\includegraphics[width=1\linewidth]{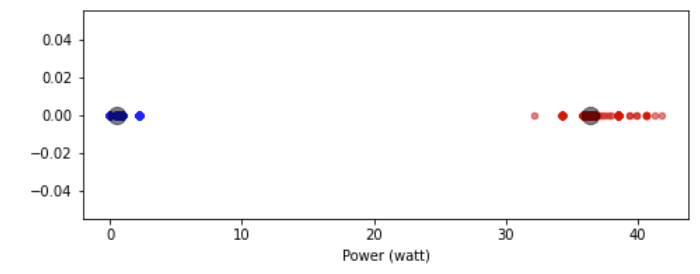}}
	\caption{Display device data}
	\label{fig:screen}
\end{figure}

\subsection{Devices}

In this work, 8 different devices are scheduled: a coffee machine, a fridge, a laptop, a microwave, a printer, a display, a thin client, and a battery based energy storage device. Two power states for each controllable device are considered. These power states are defined on the basis of the historical power data of the devices. The historical data was collected over a period of eight months, with measurements taken every 10 seconds. The K-means clustering algorithm~\cite{macqueen1967some} is applied to specify the boundary between different power states based on the power values and the density of the points. 

As an example, Figure~\ref{fig:screenHistogram} shows the historical power data and its histogram for the display device. As is seen from the histogram, as well as examined by the K-means inertia plot and elbow method~\cite{ng2012clustering}, the optimum number of clusters for the data set is two. Figure~\ref{fig:screenClustering} shows the clustering result highlighting the centroids of the clusters. The boundary between the two clusters is placed between the two centroids. Table~\ref{tbl:clustering_results} shows the cluster centroids for different devices with K-means clustering algorithm. S0 and S1 in this table, stand for two different power states.

\def\arraystretch{1.1}
\begin{table}[b]
	\caption{Centroids of the power states for different devices}
	\centering
	\setlength{\tabcolsep}{3pt}
	\begin{tabular}{r|c|c}
		 \textbf{Device} &	\textbf{Power state S0 (watt)}	& \textbf{Power state S1 (watt)} \\
%		\hline
%		\textbf{Boiler}	& 1.21 &	2109.44 \\
		\hline
		Coffee Machine &	22.48	& 2555.84 \\
		\hline
		Fridge &	0.05	& 50.8 \\
		\hline
		%		\textbf{Laptop} &	1.48e-12	& 1.95  \\
		%		\hline
%		\textbf{Laptop 1} &	1.89	& 21.68 \\ % Base Station
%		\hline
		Laptop	& 14.12 &	76.61 \\ % Projector
		\hline
		Microwave	& 4.3 &	1567.71 \\
		\hline
		Printer &	11.13 &	654.59 \\
		\hline
		%		\textbf{Reserve-568} &	5.78e-05 &	2.21 \\
		%		\hline
		%		\textbf{Reserve-566} &	0.02 &	2.12 \\
		%		\hline
		Display & 0.63	& 36.36 \\
		\hline
		Thin Client	& 0.07 &	10.52 \\
	\end{tabular}
	\label{tbl:clustering_results}
\end{table}

Each device needs to have a policy assigned to it in order for the scheduler to understand the constraints under which to schedule it. An overview of each device and the assigned policy is shown in Table~\ref{tbl:device_polices}. The coffee machine and the fridge require the \textit{repeat} policy, as these devices need to be repeatedly turned on to maintain their temperature. The laptop uses the \textit{total} policy as it needs to be charged for a total amount of time. The microwave follows a predefined \textit{pattern} based on the usage of this device by the users; it is used more during lunch time. The printer has multiple printing jobs that need to be performed, therefore it has a \textit{multiple} policy. The display and its associated thin client are only on during office hours, and therefore require a \textit{strict} policy. From the perspective of the scheduler, the battery energy storage is just another device that can be scheduled, though this device does not only consume energy, but can also return energy at a later point in time. The energy storage is assigned the \textit{battery} policy.

The total daily energy demand for these devices is approximately 15.6 kWh. The majority of this demand originates from the coffee machine, the printer, and the microwave. The coffee machine needs to repeatedly reheat water to maintain the required temperature, the printer has numerous large printing jobs that need to be completed, and the microwave is used heavily during lunch breaks.

\begin{table}[!t]
	\caption{Device Policies}
	\centering
	\begin{tabular}{c|c}
		\textbf{Device} & \textbf{Policy} \\
		\hline
		Coffee Machine & REPEAT \\
		\hline
		Fridge & REPEAT \\
		\hline
		Laptop & TOTAL \\
		\hline
		Microwave & PATTERN \\
		\hline
		Printer & MULTIPLE \\
		\hline
		Display & STRICT \\
		\hline
		Thin Client & STRICT \\
		\hline
		Battery Energy Storage & BATTERY \\
	\end{tabular}
	\label{tbl:device_polices}
\end{table}

\subsection{Energy Sources and Storage}

For local energy generation, the building has a single micro wind turbine, and a photovoltaic panel installation. The building also has battery storage. The geographical location of the building affects the energy sources, as the renewable energy generation is influenced by the weather, and the prosumers are constructed based on the day-ahead energy of the energy market. We choose as location Stuttgart, Germany.

The wind turbine is modelled after a Sonkyo Energy Windspot 1.5, for which the technical specifications are shown in Table~\ref{tbl:windspot}. The power coefficient is derived from the specifications as the manufacturer does not provide it explicitly. It should be noted that typically the specifications are optimistic, and that in practice the power coefficient will be lower. However, for the purpose of this evaluation the differences between theoretical and practical performance are not of major importance. The cost of locally generated wind energy is fixed at 0.08 €/kWh.

\begin{table}[!b]	
	\caption{Windspot 1.5 - Specifications}
	\centering
	\begin{tabular}{c|c}
		Rotor Swept Area & 12.88 m² \\
		\hline
		Rated Power & 1.5 kW \\
		\hline
		Rated Speed & 12 m/s  \\
		\hline
		Cut In Speed & 3 m/s \\
		\hline
		Survival Speed & 60 m/s \\
		\hline
		Power Coefficient & 0.11
	\end{tabular}
	\label{tbl:windspot}
\end{table}

For the photovoltaic panels, it is assumed that the building is equipped with two-axis maximum power point tracking solar panel installation. The specifications of the panels, as shown in Table~\ref{tbl:pv}, are based on the HiTech Solar 250Wp Black 60 cells panel. The installation consists of 6 panels, with a total of 1.5 kW peak power, and an area of 9.9 m². The cost is set to 0.06 €/kWh. 
% 0.15*10*1000(1-0.005(25-25))?

\begin{table}[!h]
	\caption{HiTech Solar - Specifications}
	\centering
	\begin{tabular}{c|c}
		Dimensions & 1x1.65 m \\
		\hline
		Panel Area & 1.65 m² \\
		\hline
		Cells & 60 \\
		\hline
		Technology & Monocrystalline  \\
		\hline
		Maximum Power & 250 Wp \\
		\hline
		Panel Efficiency & 15.30 \%
	\end{tabular}
	\label{tbl:pv}
\end{table}

As for the battery-based energy storage, the assumption is that the building has a single battery storage installation. The assumed energy storage is based on AlphaESS SMILE3. The technical specifications of the battery energy storage are given in Table~\ref{tbl:battery}. 

\begin{table}[!h]
	\caption{SMILE3 - Specifications}
	\centering
	\begin{tabular}{c|c}
		Capacity & 2.8 kWh \\
		\hline
		Charging/Discharging Current & 60 A \\
		\hline
		Charging/Discharging Power & 3000 W  \\
		\hline
		Depth of Discharge & 95 \% \\
	\end{tabular}
	\label{tbl:battery}
\end{table}

Both the prosumers and the grid are modelled using the day ahead prices of the German electricity market, as published on the ENTSO-E Transparency Platform. To model the grid, the hourly cost of energy is to be equal to the hourly day-ahead prices as obtained from ENTSO-E. For the purpose of this evaluation, we assume that the grid is always able to satisfy the energy demand of the building, in case the renewables and prosumers are not able to. The obtained grid prices are normalised between €0.40/kWh and €0.60/kWh. The prosumers are also modelled based on the hourly day-ahead prices. It is assumed there are 10 different prosumers available at all time. To introduce some variability, the hourly energy cost for each prosumer is sampled from Equation~\ref{eq:gaussiandiss}, where $\mu = \frac{grid\_price}{1.5}$ and $\sigma = 0.025$. The available energy for each prosumer is sampled from Equation~\ref{eq:uniformdis}, where $a = 0$ and $b = 1$.

\subsection{Results: Economic}

\begin{figure*}[ht]
	\centering
	\subfloat[Case A, renewable energy production \label{fig:case_a_renewables}]{\includegraphics[width=0.45\linewidth]{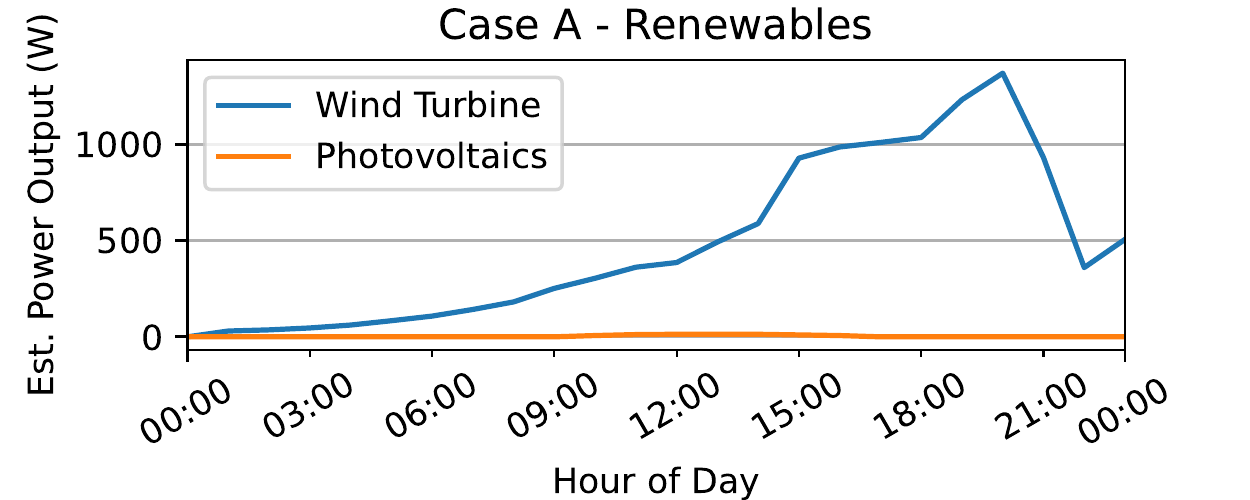}}
	\subfloat[Case A, grid and prosumers price signal \label{fig:case_a_cost}]{\includegraphics[width=0.45\linewidth]{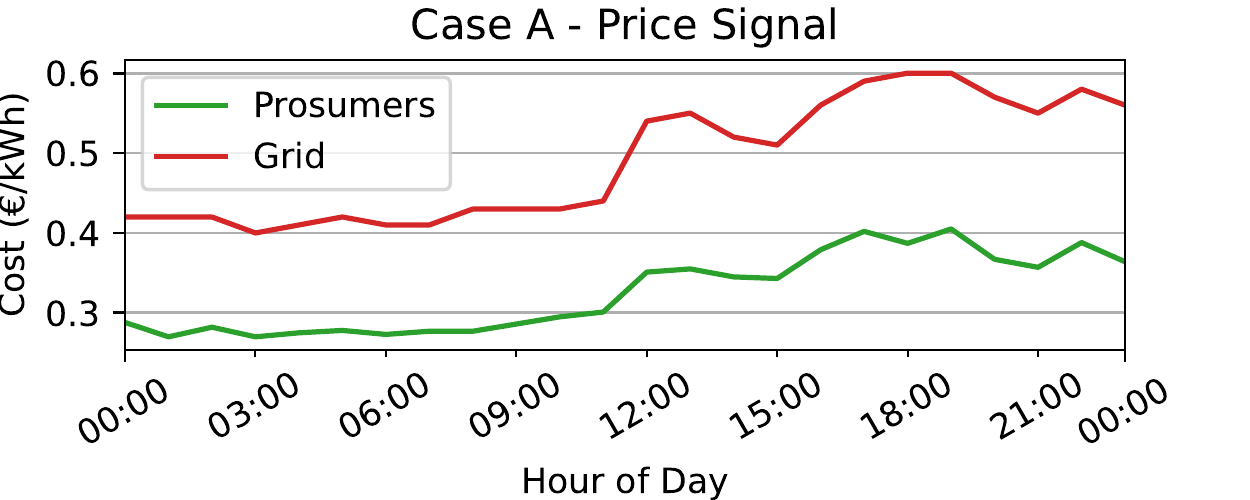}}
	\quad
	\subfloat[Case B, renewable energy production \label{fig:case_b_renewables}]{\includegraphics[width=0.45\linewidth]{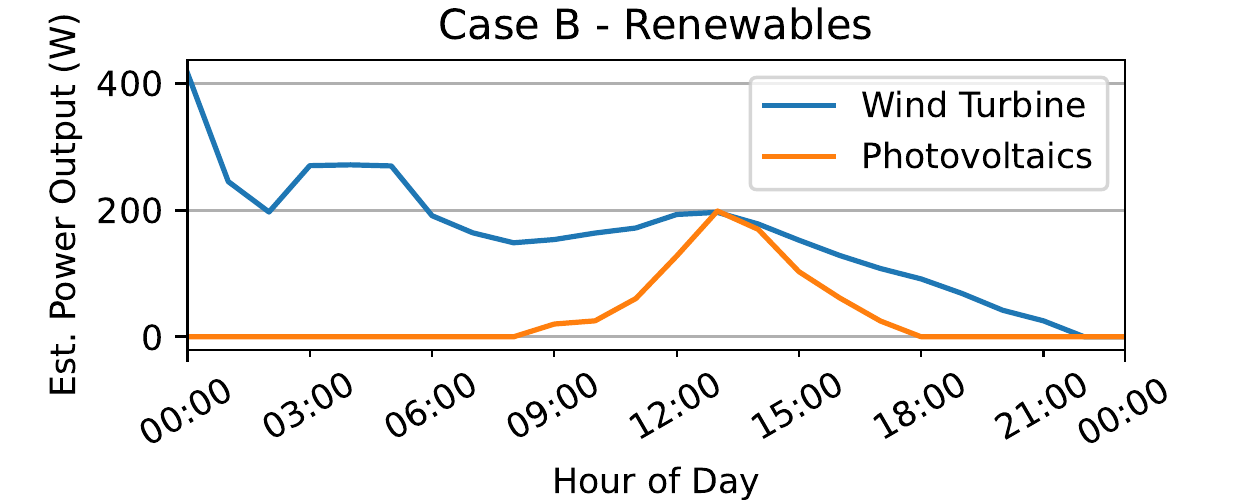}}
	\subfloat[Case B, grid and prosumers price signal \label{fig:case_b_cost}]{\includegraphics[width=0.45\linewidth]{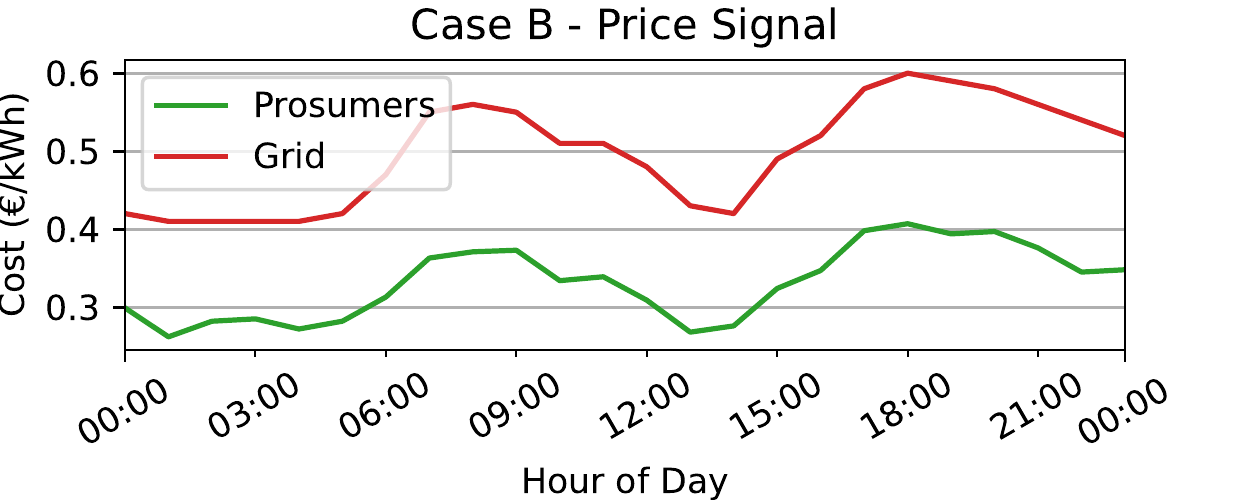}}
	\quad
	\subfloat[Case C, renewable energy production \label{fig:case_c_renewables}]{\includegraphics[width=0.45\linewidth]{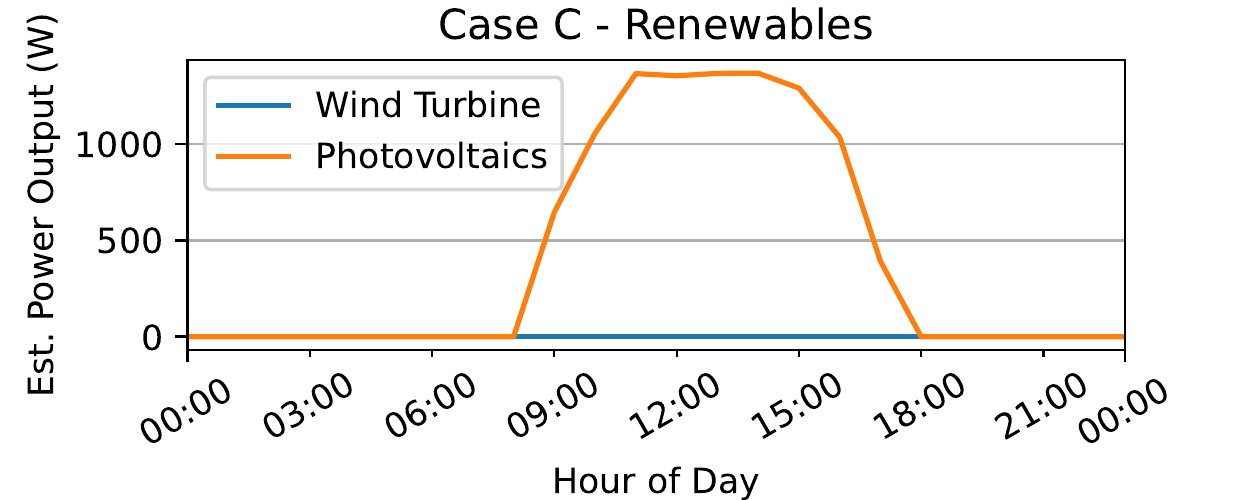}}
	\subfloat[Case C, grid and prosumers price signal \label{fig:case_c_cost}]{\includegraphics[width=0.45\linewidth]{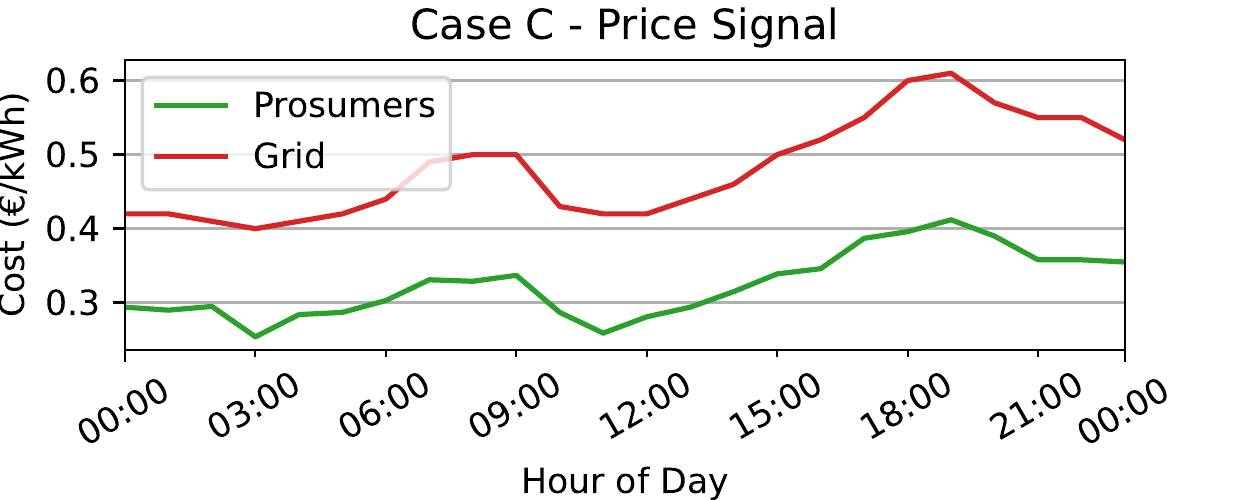}}
	\caption{The renewable energy production and price signals for each of the three considered cases}
\end{figure*}

Three distinct cases are considered in order to evaluate the economic benefits of scheduling devices with and without energy storage in the presence of locally generated renewable energy. Each case has a unique price signal, modelled on the day ahead prices of the German electricity market. Furthermore, each case also has a unique renewable energy profile which is dependent on the weather forecast for that day. The renewable energy profile and the price signal for Case A (February 5th, 2022) are shown in Figures~\ref{fig:case_a_renewables} and~\ref{fig:case_a_cost}, respectively. Case A is a cloudy and extremely windy day, especially after 14:00. The price signal for case A shows that the price is relatively low until around 11:00. The prosumer signal is the average price signal of the 10 available prosumers. Case B (February 6th, 2022), as shown in Figures~\ref{fig:case_b_renewables} and~\ref{fig:case_b_cost}, is a windy day with some PV generation throughout the day. And finally, Case C (February 8th, 2022) is a sunny day with wind speeds below the cut-in wind speed of the turbine, as can be seen in Figure~\ref{fig:case_c_renewables}. The price signal is low for the majority of the day, until approximately 16:00, as shown in Figure~\ref{fig:case_c_cost}.

% with/without battery, with/without renewables

The energy sources and their prices, as well as the list of devices, form the input of the scheduler. In total, six schedules are generated. Two schedules for each of the three cases, one with energy storage, and one without energy storage. The values of energy sources do not change for each individual case, the only change is the inclusion and exclusion of the energy storage.  The scheduling horizon is set to 24 hours, the length of each scheduling period is set to 15 minutes, and the energy storage is empty at the start of the day.

\begin{figure*}[ht]
	\centering
	\subfloat[Case A, cummulative charge \label{fig:case_a_charge}]{\includegraphics[width=0.33\linewidth]{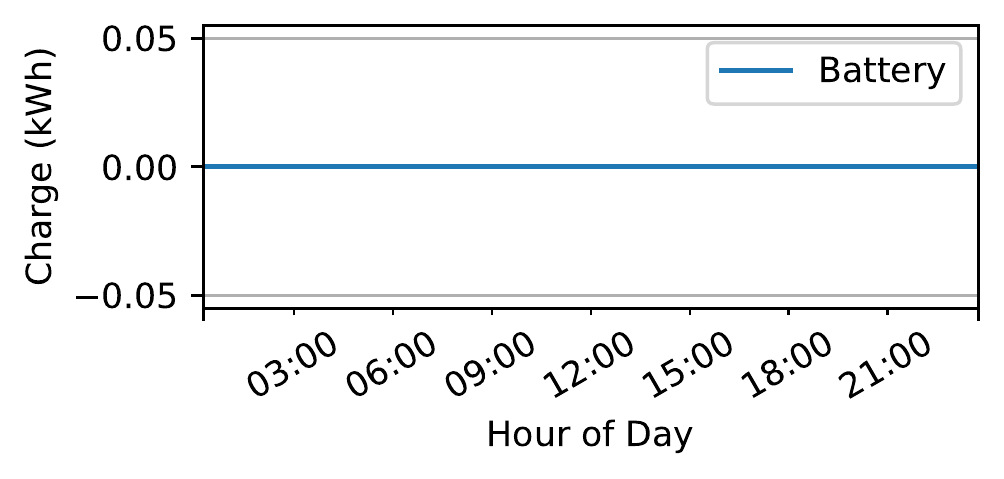}}
	\subfloat[Case A, cummulative cost \label{fig:case_a_total_cost}]{\includegraphics[width=0.32\linewidth]{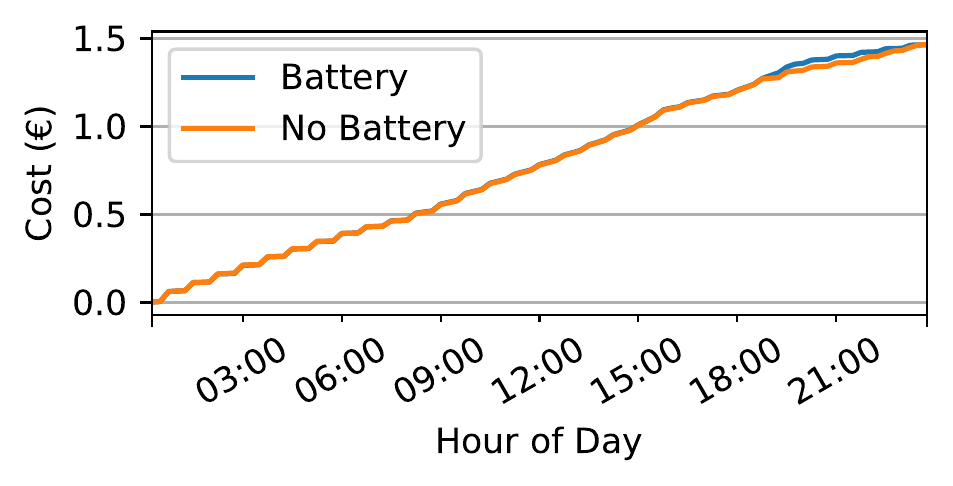}}
	\subfloat[Case A, cost per period \label{fig:case_a_total_energy}]{\includegraphics[width=0.32\linewidth]{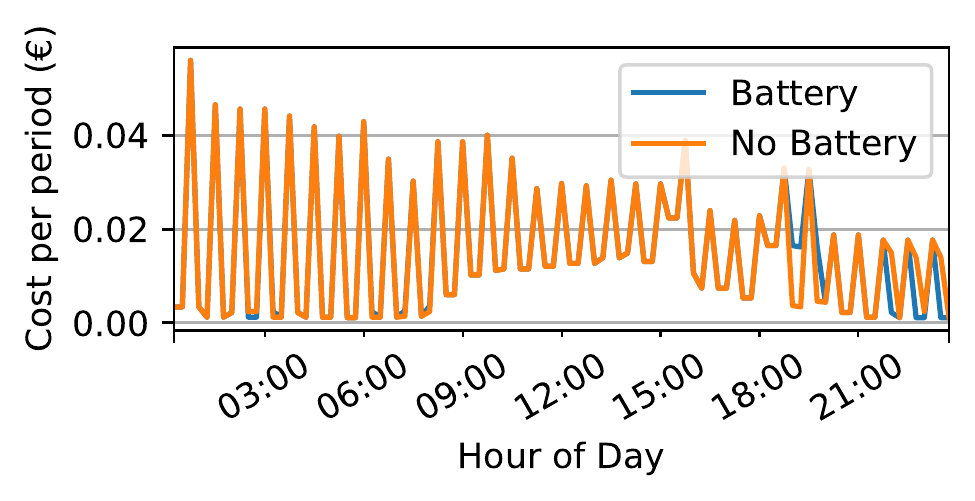}}
	\quad
	\subfloat[Case B, cummulative charge \label{fig:case_b_charge}]{\includegraphics[width=0.33\linewidth]{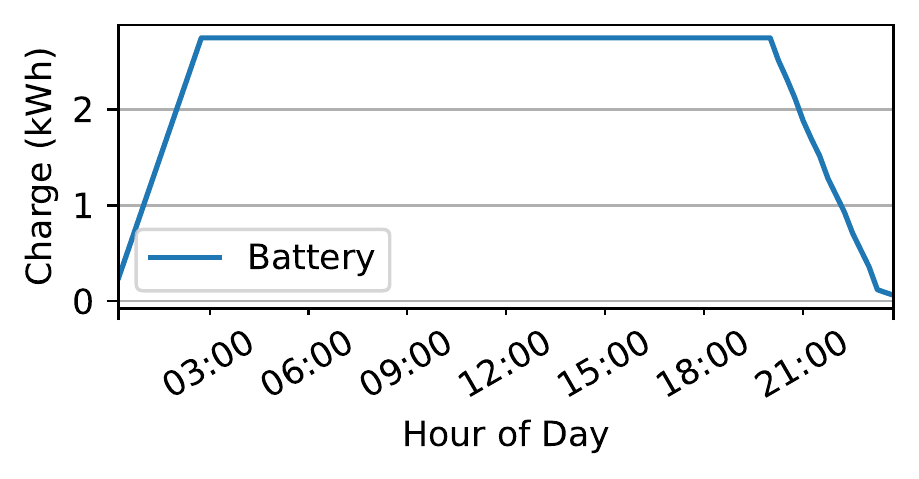}}
	\subfloat[Case B, cummulative cost \label{fig:case_b_total_cost}]{\includegraphics[width=0.32\linewidth]{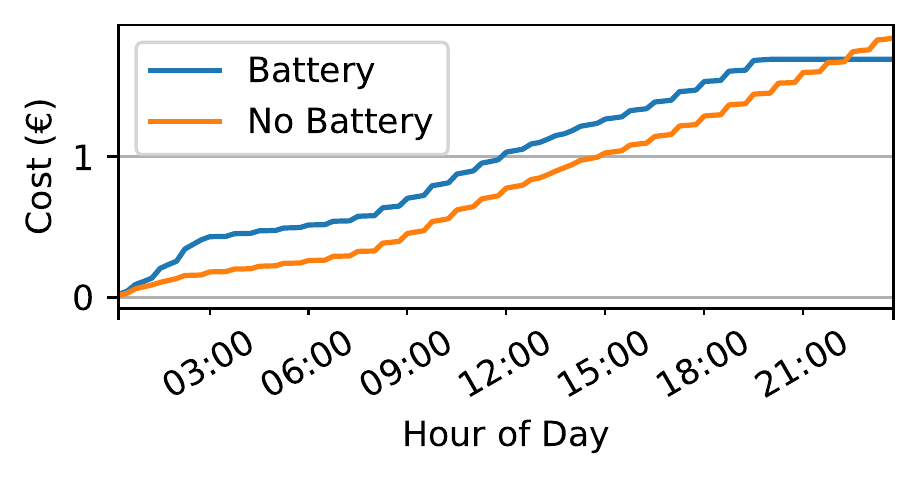}}
	\subfloat[Case B, cost per period \label{fig:case_b_total_energy}]{\includegraphics[width=0.32\linewidth]{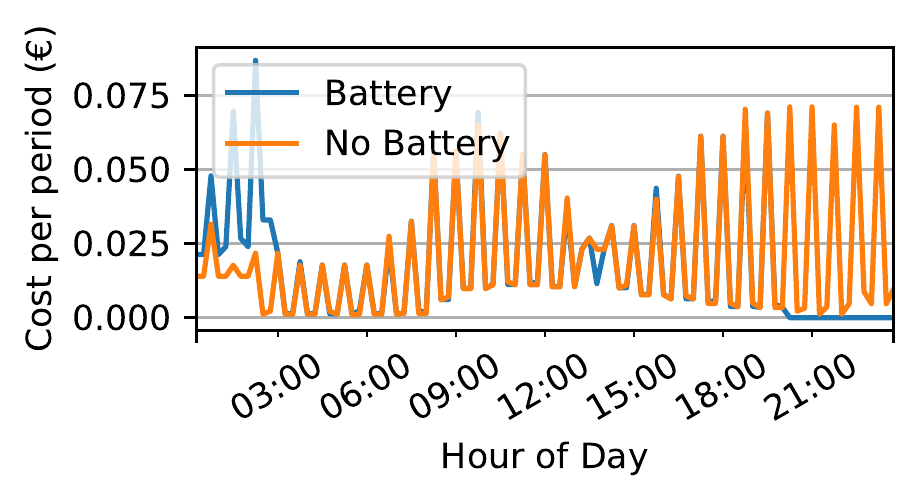}}
	\quad
	\subfloat[Case C, cummulative charge \label{fig:case_c_charge}]{\includegraphics[width=0.33\linewidth]{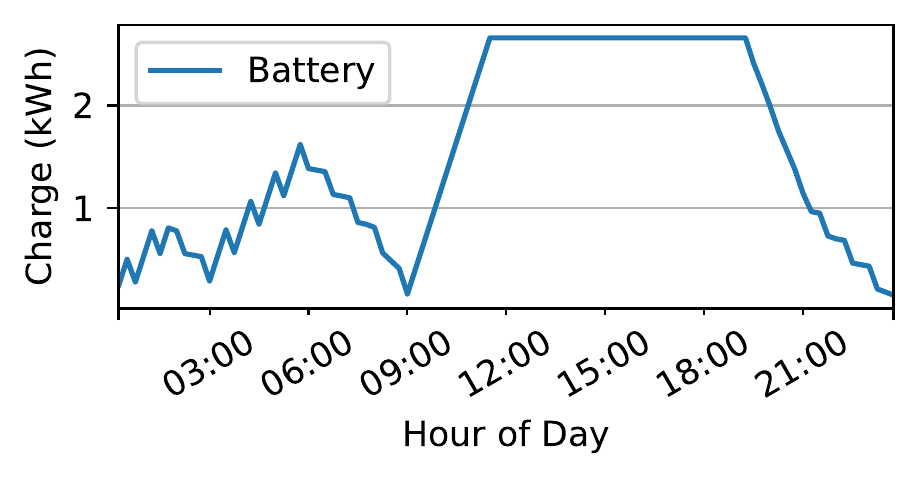}}
	\subfloat[Case C, cummulative cost \label{fig:case_c_total_cost}]{\includegraphics[width=0.32\linewidth]{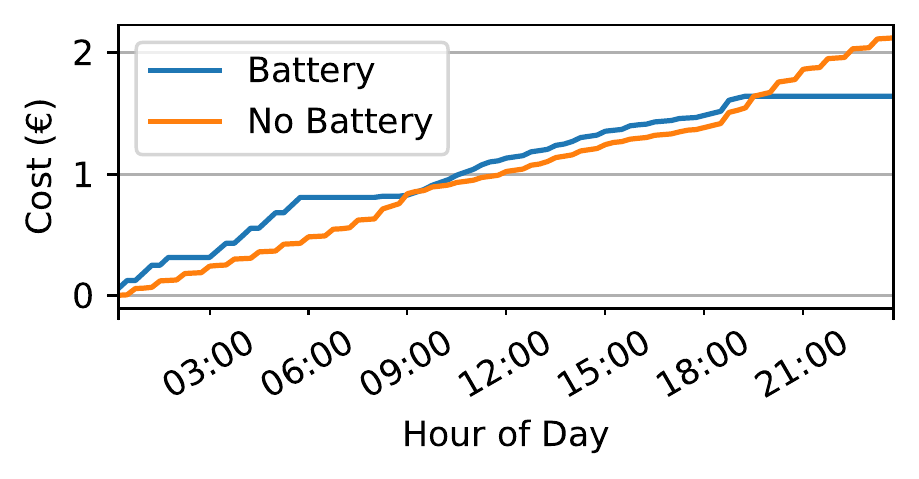}}
	\subfloat[Case C, cost per period \label{fig:case_c_total_energy}]{\includegraphics[width=0.32\linewidth]{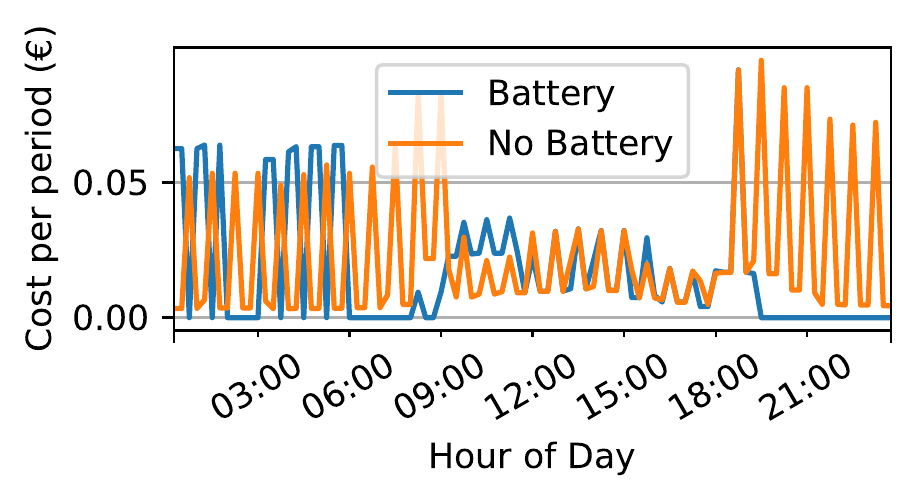}}
	\quad
	\caption{The cumulative battery charge, cumulative cost, and cost per period associated with the generated schedules}
\end{figure*}

Figure~\ref{fig:case_a_charge},~\ref{fig:case_a_total_cost}, and~\ref{fig:case_a_total_energy} show the cumulative charge of the energy storage device (battery), the cumulative cost, and the cost per period, for case A. The cumulative charge is only applicable to the schedule with energy storage. The remaining two figures visualise the results for both the schedule with energy storage, and the schedule without energy storage. It is apparent from Figure \ref{fig:case_a_charge} that the optimal schedule, i.e. the schedule with the lowest cost, does not charge or discharge the battery. The reason for this behaviour is the availability of cheap locally generated renewable energy (wind) throughout the day. The final cumulative cost is identical for both schedules, though the scheduling of devices is slightly different as can be seen in Figure~\ref{fig:case_a_total_energy}, where the cost per period differs slightly from 18:00 onwards. This difference can be explained by the parallel nature of the algorithm used by the scheduler, it is possible that the search space is explored in a different order, though this clearly does not affect the goal of minimizing total cost. The peaks in the cost per period graph are caused by the fridge and coffee machine repeatedly turning on and off with a certain periodicity.

The schedules for case B clearly exhibit different behaviour when compared to case A. Figure~\ref{fig:case_b_charge} demonstrates this, the battery is charged in the beginning of the day, and is not discharged until the end of the day. When looking at the availability of locally generated renewable energy, and the price signal, it becomes clear why case B behaves in such way. Until approximately 19:00 there is sufficient renewable energy available to meet most of the demand. Simultaneously, the price signal remains high after 17:00. Therefore, the battery is charged in the early hours of the day, as cheap locally generated renewable energy is available, and is discharged after 19:00, as the renewable energy generation is not sufficient, and the price signal as high.

For case C, as visible in Figure~\ref{fig:case_c_charge}, the charge of the battery fluctuates between 00:00 and 09:00, after which it is fully charged, and remains fully charged until 19:00. From Figure~\ref{fig:case_c_total_cost} one notices that the cumulative costs of the schedule which includes the battery are significantly lower. These savings are obtained after 19:00, when the price signal is high and there is no renewable energy available, as visible in Figure~\ref{fig:case_c_total_energy}.

An overview of the economic costs associated to each of the 6 schedules is shown in Table~\ref{tbl:cost}. The price signal and availability of locally generated renewable energy greatly affect the potential economic benefits of utilizing energy storage. Regardless of these factors, the scheduler is always able to find the optimal schedule, whether this includes utilizing energy storage or not.

\begin{table}[b]
	\caption{Economic Benefits Overview}
	\centering
	\begin{tabular}{c|c|c|c}
		Case & Battery? & Cost (€) & Cost Difference \\
		\hline
		\hline
		A & No & 1.46 & - \\
		\hline
		A & Yes & 1.46 & - \\
		\hline
		\hline
		B & No & 1.84 & - \\
		\hline
		B & Yes & 1.69 & -€0.15 (8.15\%) \\
		\hline
		\hline
		C & No & 2.12 & - \\
		\hline
		C & Yes & 1.64 & -€0.48 (22.64\%) \\
	\end{tabular}
	\label{tbl:cost}
\end{table}

\subsection{Results: Performance}

When optimizing energy it is important that the system is fast enough to be used in the building and that it does not consume more resources than it saves. 
To study this we perform the analysis on a Dell PowerEdge R7425 server with two AMD EPYC 7551 2.0GHz 32-core processors, 512GB of RAM, and read-optimized SSDs with a throughput of 560MB/s. The bottleneck of this test bed is the relatively low clock speed of the AMD EPYC 7551, which limits the number of operations per second when executing the uniform grid search.

To evaluate the performance of the algorithms, a set of test cases is generated by varying numerous parameters. These parameters include: the number of devices, scheduling horizon, number of threads, number of iterations, and algorithm variation. The number of devices ranges from 1 to 12. The scheduling horizon ranges from 2 hours to 24 hours, in 1 hour increments. The number of threads ranges from 1 to 16. The number of iterations is set to 10, and indicates how many times the same test case is run. And finally, the algorithm variations include the original algorithm from our previous work~\cite{Georgievski2012}, a variation of the original algorithm that minimizes memory operations, and the parallel algorithm proposed in this work. The original algorithm was modified to support battery device policy and the multitude of microservices. The result is 27.600 distinct test cases.

\begin{figure*}[ht]
	\centering
	\subfloat[8 devices, 24 hours\label{fig:8d24h}]{\includegraphics[width=0.4\linewidth]{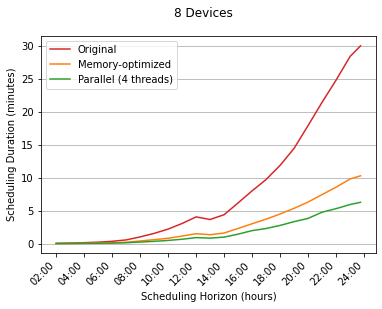}}
	\subfloat[8 devices, 24 hours, excluding original\label{fig:8d24hpo}]{\includegraphics[width=0.4\linewidth]{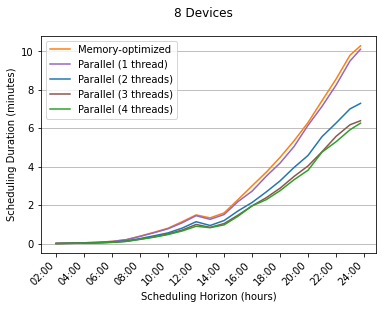}}
	\quad
	\subfloat[4 devices, 6 hours\label{fig:4d6h}]{\includegraphics[width=0.4\linewidth]{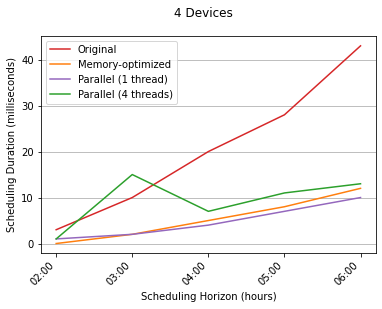}}
	\subfloat[1 to 8 devices, 24 hours\label{fig:1-8d24h}]{\includegraphics[width=0.4\linewidth]{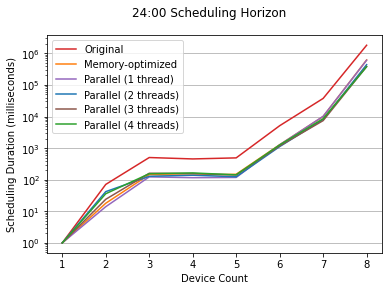}}
	\caption{Performance of the variations of the uniform grid search algorithm}
\end{figure*}

Due to the high dimensionality of the resulting data, it is not possible to visualize all of the results in their entirety. Therefore, we focus on a number of specific cases which characterize the general performance of the approach. Figure~\ref{fig:8d24h} shows the mean performance when scheduling 8 devices with a varying scheduling horizon going from 2 to 24 hours. This case is selected because the scheduling problem is reasonably complex. The time it takes to find the optimal schedule (scheduling duration) is measured in minutes. It is immediately clear that the memory-optimized and parallel algorithms perform considerably better than the original algorithm. The more complex the search space (the broader the scheduling horizon), the more pronounced the difference becomes. At the 24 hour scheduling horizon, the original algorithm (30 minutes) is 2.9 times slower compared to memory-optimized (10.2 minutes), and 4.7 times slower compared to parallel (6.2 minutes) with 4 threads.

 %remove title from last image

Figure~\ref{fig:8d24hpo} shows the same results as the previous figure, except the original algorithm is excluded, and the results for the varying thread count parameters are added. This highlights the performance differences between number of threads. The memory-optimized and parallel (1 thread) algorithms have identical performance (10.2 minutes at 24:00 hours scheduling horizon), as is expected due to the fact that the parallel algorithm is derived from the memory-optimized one. Increasing the thread count from 1 to 2, improves the performance by a factor of 1.4 (7.2 minutes at 24:00). When increasing the thread count to 3 or 4, the performance improves by a factor of 1.6 (6.2 minutes at 24:00) for both cases when compared to the memory-optimized results. 
This is due to the implementation of the concurrent priority queue. As this implementation requires synchronization when adding and removing nodes to the queue, acquiring the lock becomes the bottleneck. 

In Figure~\ref{fig:4d6h}, the case of 4 devices and a 06:00 hour scheduling horizon is considered. This case is selected because complexity-wise it is the opposite of the previous one: the search space is small. When the search space is small, the overhead of synchronizing 4 threads becomes larger than the time it takes for the single threaded variations of the algorithm to find the solution. This is clearly visible at 03:00, where the parallel algorithms (with 4 threads) requires 15 milliseconds to complete, while parallel (with 1 thread) requires only 2 milliseconds. In practice, the search space is typically more complex, and a difference of 13 milliseconds is negligible in the broader perspective of the system's execution time.

In the three preceding cases the number of devices is fixed. Figure~\ref{fig:1-8d24h} (logarithmic y-axis) demonstrates the effect of increasing the device count, and therefore increasing the search space. The complexity does not change significantly when adding the 4th and 5th devices, this is due to the policy that is assigned to these devices. Some policies (such as \textit{pattern}, \textit{strict}, and \textit{idle}) only minimally increase the search space. When ignoring the 4th and 5th devices, as the number of devices increases, the scheduling duration increases exponentially.

\begin{figure*}[ht]
	\centering
	\subfloat[8 devices, 24 hours\label{fig:6d24h16t}]{\includegraphics[width=0.47\linewidth]{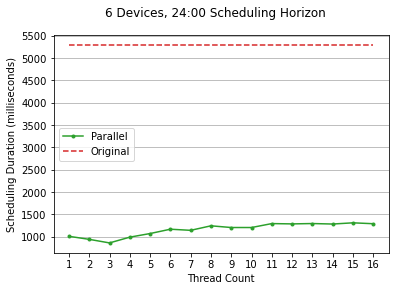}}
	\subfloat[8 devices, 24 hours, parallel algorithm only\label{fig:6d24h15terr}]{\includegraphics[width=0.47\linewidth]{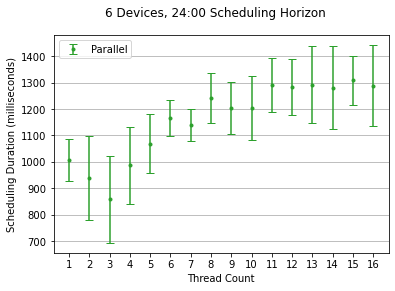}}
	\caption{Thread counts vs.\ performance}
\end{figure*}

To clearly illustrate the bottleneck problem caused by the locking mechanism of the queue, the number of devices is reduced to 6, the thread is set to range from 1 to 16, and the scheduling horizon is fixed at 24:00 hours. Figure~\ref{fig:6d24h16t} visualizes the mean scheduling duration in milliseconds with varying thread counts. The original algorithm (red dashed line) is included for reference. For this particular scheduling problem, it is clear that 3 threads provides the fastest scheduling duration. As the thread count grows beyond 3, the synchronization mechanism in the queue becomes the bottleneck. In addition, Figure~\ref{fig:6d24h15terr} shows the standard deviation for each thread count obtained after 10 iterations. 
\section{Discussion, Limitations \& Outlook}
\label{sec:discuss}

In our previous work we demonstrated the concept of scheduling devices against a price signal in order to decrease energy cost in an office building~\cite{Georgievski2012}. The present work expands that concept and improves upon it in multiple ways. First of all, by adding real time modelling of renewable energy generation, as well as the inclusion of energy storage. Second, the modelling of energy sources is supported by a loosely coupled, modular microservice architecture which enables the system to adapt to different cloud service providers. Furthermore, real world data is used for the scheduled devices, and for the prediction of renewable energy generation. And finally, significant performance enhancements are made by introducing a parallel uniform cost-search algorithm.

From the economic perspective, the inclusion of energy storage coupled with device scheduling can potentially lead to significant cost savings. However, as shown by the presented results, this depends greatly on the price signal and on the availability of locally generated renewable energy. For example, when sufficient amounts of energy are generated throughout the day, the battery is not scheduled to charge, as charging is not free and the number of charge cycles of a battery is limited. Therefore, in the worst case, the economic benefit of adding energy storage is null on a daily basis. Nevertheless, in several cases conditions are beneficial for energy storage, for example when price signals fluctuate and the locally generated renewable energy is not sufficient to meet the demand, then there are significant cost savings to be made. The results show cost savings of up to 8.15\% per day, going as high as 22.64\% under the more favourable conditions. While the evaluation focuses on the economic perspective, the algorithm is agnostic with regards to the costs. In other words, instead of providing a list of energy sources and their cost expressed in euros per kWh, one could provide instead the cost expressed as grams of CO$_2$ equivalent emissions per power unit (gCO2eq per kWh) to improve the sustainability rather then the economic costs. 

From the performance perspective, the optimization and parallelization of the uniform cost-search algorithm result in significant performance gains. The original algorithm requires up to 30 minutes to schedule 8 devices with a scheduling horizon of 24 hours. The optimizations alone reduce this time to 10 minutes. Further gains are made by parallelizing the uniform-cost search algorithm, completing the scheduling in 6.2 minutes with 4 threads. It is also possible to spread the load over multiple machines using distributed agents, where each agent solves a part of the problem. However, accessing the shared resources requires synchronization, and therefore needs to be executed sequentially. This sequentially executed section of the algorithm becomes the bottleneck. The bottleneck is a manifestation of Amdahl's law~\cite{Amdahl2007}, which in the context of parallel programming describes that the maximum possible speed-up is disproportionately limited by the sequential section of the application.

Finding the optimal solution to a scheduling problem is a time consuming task. In the present work, the scheduler operates on 15 minute intervals with a maximum scheduling horizon of 24 hours. Decreasing the size of the intervals, expanding the scheduling horizon, and increasing the number of scheduled devices all exponentially increase the complexity of the problem. To counteract these limitations, the scheduling problem can be divided into smaller problems, which are solved individually and later combined into one schedule. While these local schedules are optimal, the global schedule is likely not to be. 

The energy consumption associated with the schedule generation is not included in the economic evaluation. As is shown in the performance evaluation, the time it takes to solve the scheduling problems is in the order of several minutes. The system only needs to be available during this time period. The execution of the schedule can be performed by a simple embedded device. As per the performance evaluation, the scheduling of 8 devices requires at most 6.2 minutes when using 4 parallel threads. The 32-core AMD EPYC 7551 CPU is responsible for the majority of the system's power consumption. Assuming each of the 4 parallel threads utilises 100\% of their allocated CPU core, the overall CPU utilization is 12.5\%. According to the SPECpower\_ssj2008 benchmarks~\cite{specpower}, this would equate to a power consumption of 100W. Therefore, generating the schedule requires 0.01 kWh. We can conclude that the cost of the energy consumption of the system itself is insignificant with respect to the overall economic savings.

From a system architectural point of view, we proposed a microservice architecture that encapsulates the three critical components in several microservices. This enables the system to leverage the real time data provided by weather services and the ENTSO-E transparency platform to generate realistic price signals on the fly. Furthermore, a performance-oriented version of the uniform cost-search algorithm is proposed, which parallelizes the search for the optimal schedule. The performance evaluation shows that this parallel algorithm decreases the time to find the optimal schedule by a factor of 4.7 compared to the previous approach. And finally, this work evaluates the economic benefits of including local energy storage in the scheduling problem, demonstrating that the inclusion of energy storage can further reduce costs by up to 22.64\%, depending on the price signal and availability of renewable energy.

\section*{Acknowledgements}
We thank Alexander Lazovik, Viktoriya Degeler, and Andrea Pagani for discussion about the approach and architecture presented here as an extension of our joint previous work~\cite{Georgievski2012}.

\bibliographystyle{IEEEtran}
\bibliography{article}

\end{document}